\documentclass[mathleft
]{an}
\usepackage{graphicx}
\usepackage{times}
\begin{document}

% The following seven commands are intended for editorial usage and should be ignored by
% the author(s).

\title{ Isolated dSph galaxy KKs3 in the local  Hubble flow}

\author{I.D.~Karachentsev\inst{1}
\fnmsep\thanks{Corresponding author:
  \email{ikar@sao.ru}\newline}
\and A.Yu.~Kniazev\inst{2,3,4}\fnmsep\thanks{Based on observations made with the Southern African Large Telescope (SALT) under program
2014-2-MLT-001 (PI: Kniazev).}
\and M.E.~Sharina\inst{1}}

\titlerunning{Isolated dSph galaxy KKs3}
\authorrunning{I.D. Karachentsev et al.}
\institute{Special  Astrophysical  Observatory,  Russian  Academy  of  Sciences, Russia
\and South African Astronomical Observatory, PO Box 9, 7935 Observatory, Cape Town, South Africa
\and Southern African Large Telescope Foundation, PO Box 9, 7935 Observatory, Cape Town, South Africa
\and Sternberg Astronomical Institute, Lomonosov Moscow State University, Moscow 119991, Russia}

\received{12 June, 2015}
\accepted{31 July, 2015}
\publonline{later}


\abstract{ We present the SALT spectroscopy of a globular cluster
in the center of the nearby isolated dSph galaxy KKs3 situated at
a distance of 2.12 Mpc. Its heliocentric radial velocity is
$316\pm7$ km s$^{-1}$ that corresponds to $V_{LG} = 112$ km
s$^{-1}$ in the Local Group (LG) reference frame. We use its
distance and velocity along with the data on other 35 field
galaxies in the proximity of the LG to trace the local Hubble
flow. Some basic properties of the local field galaxies: their
morphology, absolute magnitudes, average surface brightnesses,
specific star formation rates, and hydrogen mass-to-stellar mass
ratios are briefly discussed. Surprisingly, the sample of the
neighboring isolated galaxies displays no signs of compression
under the influence of the expanding Local Void.}

\maketitle

\section{Introduction}
Numerous high-accuracy measurements of distances to nearby
galaxies performed over the past 15~years with the Hubble Space
Telescope showed that the regular Hubble flow starts directly
outside the boundary of the Local Group (LG). As described in
Karachentsev \& Makarov (2001), Ekholm et al. (2001), and
Karachentsev et al. (2002), the radial velocities and distances of
the local field galaxies with respect to the LG centroid comply
well with the Hubble relation $V_{LG}= H\times D_{LG}$. Along with
this, the local value of the Hubble parameter $H$ turns out
similar to its global value, $H_0=73$ km s$^{-1}$ Mpc$^{-1}$, and
the peculiar radial velocity dispersion is only
$\sigma(V_{pec})\sim30$ km s$^{-1}$. Approximately the same
peculiar velocity dispersion is also typical for the barycenters
of the nearby galaxy groups (Karachentsev, 2005). A considerable
number of papers are devoted to the interpretation of the observed
data on the ``cold'' Hubble flow (Chernin 2001, Chernin et al.
2004, Teerikorpi et al. 2005, Chernin 2008, Hoffman et al. 2008,
Peirani \& Pacheco 2008, Tikhonov \& Klypin 2009, and Chernin et
al. 2015).

With the low peculiar radial velocity dispersion, the curvature of
the local Hubble flow, caused by the retarding  gravitational
influence of the LG as a local attractor, manifests itself
distinctly. As reported by Karachentsev et al. (2009), the radius
of the zero-velocity surface around the LG is $R_0 =
(0.96\pm0.03)$ Mpc, to which the total mass of the Local Group
$M_T(LG)= (1.9\pm0.2)\times 10^{12} M_{\odot}$ corresponds in good
agreement with the sum of individual mass estimations for the
Milky Way (MW) and the Andromeda galaxy (M31).

Over the last years, the accuracy in measurements of the distances
to the nearby galaxies using the tip of the red giant branch
(TRGB) method was raised up to $\sim5$\% (Rizzi et al. 2007). The
most complete reviews of the distance estimates are presented in
the ``Extragalactic Distance Database'' (=EDD, Tully et al. 2009)
and the ``Updated Nearby Galaxy Catalog'' (=UNGC, Karachentsev et
al. 2013). Both are available at: http://edd.ifa.hawaii.edu and
http://lv.sao.ru/lvgdb; the reviews are periodically updated.

It is obvious that the isolated (field) galaxies, the velocities of
which are not contaminated by large virial motions such as found for 
group members, are
the most suitable ones for tracing the local Hubble flow. When
analyzing the Hubble pattern of the peculiar velocities
(Karachentsev et al. 2009), 30 galaxies outside the LG with
distances of 0.7 Mpc $<D_{LG}<3.0$ Mpc were considered. Since
then, only two new dwarf galaxies, Leo~P (Giovanelli et al. 2013)
and KK~258 (Karachentsev et al. 2014), have been added to that
list. One more isolated dSph galaxy KKs3 was recently discovered
at a distance of 2.12 Mpc (Karachentsev et al. 2015), although its
radial velocity was unknown. We will determine below the radial
velocity for this galaxy and consider the location of KKs3 in the
local Hubble flow.

 \section{Surface photometry of KKs3}
For surface photometry, we used the fully processed
distortion-corrected  F606W and F814W images of KKs3 
obtained with the Hubble Space Telescope Advanced Camera for Surveys
(HST/ACS, SNAP 13442, PI R.B.Tully). Foreground stars were removed from the
frames. The sky background in the ACS images was insignificant,
but to remove possible slight large scale variations, the sky was
approximated by a tilted plane created from a two-dimension
polynomial with the use of the least-squares method. The accuracy
of the sky background determination was about 2\% of the original
sky level. To measure a total galaxy magnitude in each band, a
galaxy image was first fitted with concentric ellipses. Then
integrated photometry was performed in these ellipses. The
total magnitude was estimated as an asymptotic value of the radial
growth curve. The measured total magnitude and colour turned out
$V = 14.47\pm0.05$ and $(V - I) = 0.77\pm0.03$ respectively. The
estimated errors include the photometry and sky background
uncertainties, as well as the transformation errors from
instrumental ACS magnitudes to the standard $V$ and $I$ magnitudes
(Sirianni et al. 2005).

Azimuthally averaged surface brightness profiles for KKs3 were
obtained by differentiating the galaxy growth curves with respect
to semiaxes. The resulting profiles in the $V$ and $I$ bands are
shown in Fig.~1. The surface brightness profile can be fitted by
an exponential law with an exponential scale length of $34\farcs0$ and
$31\farcs8$ in the $V$ and $I$ bands respectively. The unweighted
exponential fits the profiles with an effective radius of $57\farcs0$ and
$53\farcs3$ and effective surface brightness of 24.87 and 24.01
($^m/\sq\arcsec$)  in the $V$ and $I$ bands respectively.

Observations of KKs3 with the ACS revealed a globular cluster
located in the dwarf galaxy center. It produces a
narrow central peak in the right upper panel of Fig.~1. Our photometry 
of the globular cluster yields
its total magnitude $V = 18.70$ and color $V - I = 0.85$ within an
aperture of $1\arcsec$ radius.

\section{Spectroscopic observations of KKs3}

Spectroscopic observations of the globular cluster were conducted
with the Southern African Large Telescope
(SALT)(Buckley et al. 2006, O'Donoghue et al.
2006) on January 15--17, 2015 with the Robert Stobie Spectrograph
(RSS, Burgh et al. 2003, Kobulnicky et al. 2003). The high
resolution long-slit spectroscopy mode of the RSS was used with a
$1\farcs25$  slit width. The grating pg0900 with blocking filter
pc03400 and Camera Station parameter 26.5 was used to cover the
spectral range of 3700--6700 \AA \, with a reciprocal dispersion
of 0.97 \AA \, pixel$^{-1}$ and  spectral resolution of FWHM=5~\AA. 
The seeing during the observation (twilight time mostly) was
about 1--2\arcsec.
The total exposure time was 5200s divided into 5 subexposures 
of $\sim$1000s each in order to remove cosmic rays. A spectrum of Ne
comparison arcs was obtained to calibrate the wavelength scale.
Primary reduction of the data was done with the SALT science
pipeline (Crawford et al. 2010). The long-slit reduction was done
later, in the way described in Kniazev et al. (2008). We note that
SALT is a telescope with a variable pupil, and its illuminated
beam changes continuously during the observations. This makes the
absolute flux/magnitude calibration impossible, even using
spectrophotometric standard stars. The one-dimensional scan of the
spectrum in the range of 3700 -- 6700 \AA\, is presented in Fig.~2.
As the data suggest, there are absorption Balmer lines and $H$ and
$K$ doublet of CaII. They yield the heliocentric radial velocity
$V_h = 316\pm7$ km s$^{-1}$, or the Local Group centroid velocity
$V_{LG} = 112$ km s$^{-1}$.

\section{The Hubble flow for the local field galaxies}
The typical separation between the group centers in the Local Volume
is 3--4 Mpc. Thus, we defined the local Hubble flow to have a 
distance limit of 3.5 Mpc. According to the UNGC catalog
(Karachentsev et al. 2013), in the area of this radius there are
125 galaxies with measured radial velocities and accurate distance
estimations. Each galaxy from the UNGC is characterized by a
``tidal index''

 $$\Theta_1 =\max[\log(M^*_n/D^3_n)] - c, \,\,\, n=1,2,\ldots N, $$
where $M^*_n$ denotes the stellar mass of the nearby galaxy and
$D_n$ is its spatial distance from the galaxy under study. Among
the numerous neighbors, there is a so-called Main Disturber (=MD),
the tidal force of which, $\sim M^*/D^3$, dominates all other
neighbors. The parameter $c = -10.96$ was chosen so that the
galaxy with $\Theta_1<0$ was situated outside the ``zero-velocity
surface'' of the radius $R_0$ around its MD. By selecting the
galaxies with a tidal index of $\Theta_1<0$, we excluded, thereby,
all the members of the LG and of other nearby groups. As the value
$\Theta_1$ is derived with a certain error, we selected the
objects with $\Theta_1\leq0.2$ for our sample. The fulfilment of
this condition gives us a model sample of 36 field galaxies, the
velocities of which do not include a considerable virial component.
Table 1 shows various characteristics of these galaxies. We
emphasize that the galaxies in this Table present the most
complete sample of the nearby field galaxies, the properties of
which suffer from observational selection effects to the
minimal degree only. The characteristics of this sample can serve
as a basis to check various simulations of the
field galaxy population.

The observables of the galaxies from Table 1 were taken from the
UNGC catalog. The table columns contain: (1) the galaxy name; (2),
(3) the supergalactic coordinates; (4) the heliocentric distance
of the galaxy and its error; (5) the distance of the galaxy from
the LG barycenter on the assumption that it is located on the line
between the MW and M31 at a distance of 0.43 Mpc away from the MW
(Karachentsev et al. 2009) here; the  mass ratio for  two
dominating galaxies, M(M31) : M(MW)=5 : 4, corresponds to that
value; (6) the heliocentric radial velocity of the galaxy and its
error; (7) the radial velocity of the galaxy in the LG reference
frame with the standard apex parameters adopted in the NASA
Extragalactic Database (NED, http://ned.ipac.caltech.edu/); (8),
(9) the tidal index $\Theta_1$ and the name of the MD that exerts
the greatest tidal force on the galaxy; (10) the peculiar radial
velocity of the galaxy relative to the regression line in Fig.~3;
(11) the morphological type as per de Vaucouleurs scale; (12) the
average surface brightness of the galaxy in the $B$ band within
the Holmberg isophote of $26.5^m/\sq\arcsec$; (13) the absolute
extinction-corrected $B$ magnitude of the galaxy; (14) the
specific star formation rate which was estimated from the
ultraviolet (FUV) flux measured with the GALEX satellite (Gil de
Paz et al. 2007); (15) the neutral hydrogen mass-to-stellar mass
ratio estimated by the luminosity of the galaxy in the $K$ band at
$M^*/L_K=1\times M_{\odot}/L_{\odot}$ (Bell et al. 2003). For some
early-type galaxies $(T<0)$, only upper values of sSFR and
$M_{HI}/M^*$ are given.

Figure 3 shows the distribution of 36 nearest isolated galaxies by
distances and radial velocities in the LG rest frame. The galaxies of
late $(T>0$) and early types are denoted by the solid and open
circles respectively (with distance errors). The dashed straight
line denotes the undisturbed Hubble flow with a parameter of $H_0$
=73 km s$^{-1}$ Mpc$^{-1}$. The solid line corresponds to the
regression of radial velocity on distance for a canonic 
Lem\'{e}tre-Tolman model with the parameter $R_0=0.96$ Mpc.

The measured radial velocity of the dSph  galaxy  KKs3 fixes its
position by 45 km s$^{-1}$ lower than the regression line. The M31
galaxy exerts the greatest gravitational force on KKs3, although
their mutual separation is too large for M31 to cause the high
peculiar velocity of KKs3.

Apart from KKs3, the other new member of the local Hubble flow,
Leo~P, has a nearly zero peculiar velocity. The distance to this
dwarf irregular galaxy, $D=1.90\pm0.05$ Mpc, was measured by
L.~N.~Makarova with the TRGB method using the images of this
galaxy from the Hubble Space Telescope Archive (G0 13376, PI
K.McQuinn).

It should be also noted that the radial velocity $V_h=-135\pm2$ km
s$^{-1}$ of the dSph galaxy KKR~25 at first was incorrectly
estimated from the HI line (Huchtmeier et al. 2000) because of
confusion with the Galactic HI line. The new estimate
$V_h=-79\pm8$ km s$^{-1}$ derived from the optical spectrum
(Makarov et al. 2012) notably shifted the galaxy in the Hubble
diagram. The same happened to the transition-type dwarf galaxy
Tucana. The first estimation of its radial velocity,
$V_h=-132\pm5$ km s$^{-1}$, was derived with association of the
HI-cloud with it (Oosterloo et al. 1996). However, the subsequent
measurements of the velocity for the galaxy stellar component
increased the value by 62 km s$^{-1}$ (Fraternali et al. 2009).

In the right upper corner of the Hubble diagram, there are five
galaxies: DDO~99, DDO~125, DDO~181, UGC~8833, and DDO~190, which
neighbor with each other in the sky and are situated in front of
the nearby cloud of  dwarf galaxies, CVnI. Attraction to  the
CVnI cloud may cause additional velocity directed away from the LG.

 \section{Peculiar velocity pattern around the LG}
The apex parameters of the Sun's motion relative to the LG
centroid in Galactic coordinates:
\begin{equation}
V_A=316\pm5 \, {\rm km\, s}^{-1}, l_A=93^{\circ}\pm2, b_A=-4^{\circ}\pm2, 
\end{equation}
adopted as standard in NED, were determined by velocities and
distances of 20 nearest galaxies in the LG vicinity (Karachentsev
\& Makarov 1996). Over the last years, new galaxies have been
discovered near the LG, and radial velocities and distances for a
number of old ones have been refined. As a result, the parameters
of the standard apex could change. Analysis of the peculiar radial
velocities of nearby galaxies with respect to the standard apex
(1) showed (Karachentsev et al. 2009) that there was a small
dipole component with the parameters:
\begin{equation}
V_d=24\pm4 \,{\rm km\, s}^{-1}, l_d=325^{\circ}, b_d=-46^{\circ}.
\end{equation}
In the supergalactic coordinates, the apex of this dipole is
situated near the equator of the Local supercluster:
$SGL_d=227^{\circ}, SGB_d=1^{\circ}$.

Figure 4 shows the distribution of the peculiar velocities of the
nearby isolated galaxies from Table 1 depending on the
supergalactic longitude. The notations for the late and early type
galaxies are the same as in Fig.~3. The vertical intervals denote
the errors in estimation of the galaxy velocities caused by distance 
errors. The wavy line
corresponds to the above-mentioned dipole with an amplitude of 24
km s$^{-1}$. As seen from this pattern, peculiar velocities of the
most galaxies follow approximately the dipole wave. Five galaxies 
toward the CVnI cloud (contoured), as well as Tucana and Leo~A,
deviate from the general tendency. Both the Tucana and Leo~A are
situated close to the zero-velocity surface for the LG. They might
have the dynamic history different from the other, more distant LG neighbors.
The peculiar velocity dispersion for all the galaxies in Fig.~4 is
32 km s$^{-1}$, and it decreases to 23 km s$^{-1}$ when excluding
5 galaxies in the CVnI region. A considerable portion of this
value accounts for the dipole component and the errors in distance
estimation. If we take the dipole effect into consideration, the
corrected values of the standard apex (1) will be as follows:
\begin{equation}
 V_a=314 \, {\rm km\, s}^{-1}, l_a=91^{\circ}, b_a=-8^{\circ}. 
\end{equation}
Notice that the direction toward the closest rich cluster
Virgo $(SGL = 103^{\circ}, SGB = -2^{\circ}$) is not associated
with any characteristic feature in the pattern $V_{pec}$ vs.
$SGL$.

As is well-known, the LG and other neighboring galaxy groups are
located within the flat structure (wall) which emborders the Local
Void. Weygaert \& Schaap (2007) and Tully et al. (2008) noted that
a completely empty void in a topologically flat $\Lambda$CDM
universe expands at a rate of 16 km s$^{-1}$Mpc$^{-1}$. If a void
is not completely empty, the expansion is going at a slower rate.
This fact makes us consider the situation with local peculiar
velocities depending on the supergalactic SGB latitude. Let us
note that the Local Void occupies approximately one-fourth of the
whole sky and its center is located close to the north
supergalactic pole.

The distribution of 36 neighboring isolated galaxies by peculiar
velocities and supergalactic latitude is presented in Fig.~5. The
notations are the same as in the previous figure. As one can see,
at high latitudes $\mid SGB\mid > 30^{\circ}$ there are only 7
galaxies instead of the expected 18 with a uniform random
distribution of them in the sky. Shortage in the number of
galaxies in the supergalactic polar caps could be conditioned by the 
fact that the poles of the Local supercluster lie in the region of a
considerable Galactic extinction. However, sky surveys in the HI
line are almost insensitive to the extinction. For this reason, the
observed shortage in gas-rich dIr galaxies in the region $\mid SGB\mid >
30^{\circ}$ is probably caused by the intrinsic tendency of
such kind of objects to be concentrated toward the local
supercluster plane.

In the picture of the Local Void expansion, three galaxies,
SagdIr, DDO~210, and KKR~25 with $ SGB > +30^{\circ}$ should have
had negative peculiar velocities. With their average height above
the supergalactic plane $\langle SGZ\rangle = +0.95\pm0.12$ Mpc,
the expected effect for a completely empty void would have
amounted to about --15 km s$^{-1}$. However, the observed average
peculiar velocity is $(+16\pm11)$ km s$^{-1}$. Thus, we can not
recognize the expected effect of the Void expansion. If the wall, in
which the LG is located, moves as a whole in the direction opposite
to the Local Void, then four galaxies: HIZSS3, NGC~3109,
Sex~A, and Sex~B with $ SGB < -30^{\circ}$, should also have had
negative peculiar velocities. Although, their position in Fig.~5 does
not agree with this assumption: the average peculiar velocity of
these galaxies is $+21\pm5$ km s$^{-1}$. It is obvious that the
behavior of the neighboring isolated galaxies in the 
\{$V_{pec}, SGB$\} diagram needs to be analyzed in detail with the
involvement of reconstructing their orbits by numerical modeling
(Shaya et al. 1995).

       \section{Some properties of the local orphan galaxies}
As emphasized above, Table 1 shows the sample of the field
galaxies which is least subject to various observational selection
effects. It is reasonable to suppose that this sample is almost
100\% luminosity complete up to the absolute magnitude $M_B=-9^m$.

Figure 6 presents a mosaic of distribution of a number of the
nearest field galaxies by basic parameters: the absolute
magnitude, the average surface brightness, the specific star
formation rate, and the retio of neutral hydrogen mass to
stellar mass of the galaxy. Five galaxies of early morphological
types are highlighted in green. These data allow us to draw the
following inferences.

\begin{itemize}
\item The main population of the field galaxies consists of dwarf
galaxies with the median absolute magnitude $M_B\simeq-13.0^m$.
More than half of them are gas-rich dIr galaxies (T =10) which are
easily detected in the HI surveys.

\item The distribution of the isolated galaxies by the average
surface brightness has the peak and the median near
$24.5^m/\sq\arcsec$, which is typical for the members of groups and
clusters. It seems unlikely that among the field dIr galaxies, there is 
a hidden population of objects with a very low surface
brightness ($SB\geq 26^m/\sq\arcsec$), which remains still
undetected. However, it should be pointed out that the large population
of extremely low surface brightness dSph systems (satellites of the MW and M31) 
was found during the last 10 years only with start of the Sloan survey and systematic 
studies of M31 halo. Almost all of these ultra-faint objects are dSphs.
A population of such kind objects can also exist in the general field,
being undetected so far with any modern HI and optical surveys.

\item The field galaxy distribution by the specific star formation rate
shows the peak near $\log(sSFR)\simeq-10.1$. Note that the value
of sSFR is of the same dimension, yr$^{-1}$, as that the Hubble
parameter, $H_0$, having the quantity $\log H_0 = -10.14$. This
coincidence means that a typical representative of the field galaxies 
is able to reproduce its stellar mass over the cosmological time
$H_0^{-1}$ with the currently observed star formation rate.

\item About 70\% of the isolated galaxies are characterized by
the ratio $\log(M_{HI}/M^*)$ in the range of [--0.5, +0.5] with
the median --0.25. In other words, the majority of field galaxies
are in the middle of their evolution process in transforming the
gas component into the stellar one.

\item Among the field galaxies, there are some objects of early
types: KKR~25 (dSph), KKs3 (dSph), KK~258 (dTr), and Tucana (dTr),
in the population of which the old stars ($t > 10^{10}$ years)
prevail. These dwarf systems have low luminosity, low surface
brightness, poor gas content, and low current star formation rate.
To explane the presence of the isolated dSph galaxies (as well as 
the phenomenon of the isolated S0 galaxy NGC~404) one could invoke
a scenario of ''cosmic web stripping'' such as proposed by 
Ben\'{i}tez-Llambay et al. (2013).

\end{itemize}

    \section{Concluding remarks}
The studied Local Volume amounts to about 160 Mpc$^3$ after
subtracting the ``infall zones''around the LG and some other
neighboring galaxy groups. In this Volume, 36 field galaxies form
the average density of $\sim0.22$ gg/Mpc$^3$. Owing to the unique
proximity of the Local Volume, almost all the neighboring isolated
galaxies have been detected so far. Some incompleteness of our sample can be
evident in the lack of a systematic survey of the northern sky
(Dec.$>+40^{\circ}$) in the 21-cm line, which could supplement
two--three new Leo~P-type irregular dwarfs. Another reason for the data
incompleteness in Table 1 can be the difficulty in detecting
gas-poor spheroidal dwarf galaxies of a low surface brightness at
high supergalactic latitudes, where the optical extinction exceeds 1
mag. However, the number of the known isolated dSph galaxies in
the rest of the transparent sky region is quite small.

For all the galaxies from our sample, the radial velocities are
measured, and the distances are estimated with an accuracy of
$\sim5$\%. Every isolated galaxy from Table 1 has the estimations
of stellar mass and mass of gas, and the data on the current star
formation rate in them. The wealth and completeness of the
observed data on the nearby field galaxies make this sample
essential while checking various models of isolated galaxy
formation by means of N-body simulations.

 \acknowledgements
Spectral observations reported in this paper were obtained with the
Southern African Large Telescope (SALT).
This work is supported by the Russian Scientific Foundation grant 14--12--00965. 
A.Y.K.\ acknowledges the support from the
National Research Foundation (NRF) of South Africa.
The authors are
grateful to L.~N.~Makarova for providing the data on the distance
to Leo~P galaxy.

{} 
\clearpage
%\begin{turn}
%\begin{rotate}
%\footnotesize
\begin{table}[H]
\caption{The  Hubble flow for the nearby field galaxies}
\scriptsize
%\begin{center}
\begin{tabular}{lrrccrcrlrrrcrrr} \hline
 Galaxy  &   $SGL$  &   $SGB$  & $D_{MW}\pm$ &  $D_{LG}$ &  $V_h\pm$ & $V_{LG}$ & $\Theta_1$&  $MD$ & $V_{pec}$&  $T$ &  $SB$  &  $M_B$ &$\lg[sSFR]$& $\lg M_{HI}/M^*$\\
     &   $\circ$  &   $\circ$  &   Mpc   &  Mpc  &  km s$^{-1}$  & km s$^{-1}$ &      &     & km s$^{-1}$&    & $m/\sq\arcsec$&  mag &  yr$^{-1}$ &          \\
\hline                                                                                                                     \\
  (1)    &   (2)  &   (3)  &  (4)    & (5)   &  (6)   & (7)  &  (8) & (9) & (10)&(11)& (12) & (13) &  (14)  & (15)     \\
\hline                                                                                                                       \\
WLM      &  277.81&   8.09 & 0.97 .02&  0.82 & -122  2&  -16 &  0.0 &  M31&  1 &   9&  24.8& -14.1& - 9.93 & 0.14 \\
ESO349-31&  260.18&   0.40 & 3.21 .16&  3.10 &  221  6&  230 &  0.0 & N253&  13 &  10&  24.8& -11.9& -10.14 & 0.01 \\
NGC0055  &  256.25&  -2.36 & 2.13 .10&  2.10 &  129  2&  111 &  0.1 & N300& -12 &   8&  24.6& -18.4& - 9.70 &-0.04 \\
NGC0300  &  259.81&  -9.50 & 2.15 .10&  2.11 &  146  2&  116 &  0.1 &  N55& -8 &   7&  24.4& -17.9& - 9.97 &-0.08 \\
NGC0404  &  331.85&   6.25 & 3.05 .15&  2.63 &  -48  9&  195 & -0.8 &Maffei2&  20 &  -1&  22.4& -16.5& -11.49 &-1.35 \\
KKs3     &  222.81& -21.34 & 2.12 .07&  2.44 &  316  7&  112 & -1.4 &  M31& -45 &  -3&  25.4& -11.8& -13.64 &-2.34 \\
HIZSS003 &   33.26& -78.42 & 1.67 .17&  1.79 &  288  3&  108 & -1.1 &  M31&   17 &  10&  23.4& -12.8& -11.53 &-0.22 \\
NGC2403  &   30.80&  -8.31 & 3.18 .16&  3.15 &  133  1&  270 &  0.2 &DDO44&  48 &   6&  24.2& -19.2& - 9.95 &-0.60 \\
UGC04879 &   47.61& -15.01 & 1.36 .06&  1.34 &  -25  4&   33 & -0.7 &  M31&  -9 &   9&  24.9& -11.9& -10.27 &-1.00 \\
LeoA     &   69.91& -25.80 & 0.80 .04&  0.96 &   24  1&  -40 & -0.1 &  MW & -39 &  10&  25.3& -11.7& -10.15 & 0.11 \\
SexB     &   95.46& -39.62 & 1.43 .07&  1.69 &  300  1&  110 & -0.8 &  MW &  29 &  10&  24.3& -14.0& -10.20 &-0.12 \\
NGC3109  &  137.96& -45.10 & 1.32 .06&  1.68 &  403  2&  110 &  0.2 &Antlia&  31 &   8&  25.0& -15.7& - 9.95 & 0.20 \\
SexA     &  109.01& -40.66 & 1.43 .07&  1.74 &  324  2&   94 & -0.8 &  MW &  8 &  10&  24.2& -14.0& - 9.49 & 0.33 \\
LeoP     &   84.87& -28.90 & 1.90 .05&  2.14 &  262  2&  135 & -1.3 &  MW &   8 &  10&  24.4& - 9.3& -10.10 &-0.17 \\
NGC3741  &   67.96&  -2.08 & 3.22 .16&  3.34 &  229  2&  263 & -0.7 &  M81&  25 &   9&  23.8& -13.2& - 9.69 & 0.50 \\
DDO99    &   74.93&  -2.12 & 2.64 .14&  2.74 &  251  2&  257 & -0.8 &N4214&  72 &  10&  25.3& -13.5& - 9.71 & 0.32 \\
IC3104   &  195.83& -17.06 & 2.27 .19&  2.61 &  429  4&  170 & -1.2 &N5128&   -3 &   9&  24.2& -14.8& -10.75 &-1.48 \\
DDO125   &   72.85&   5.93 & 2.74 .14&  2.81 &  206  2&  251 & -0.8 &N4214&  60 &   9&  24.4& -14.3& -10.37 &-0.62 \\
GR8      &  102.98&   4.67 & 2.13 .11&  2.39 &  217  2&  139 & -1.4 &  MW & -13 &  10&  24.4& -12.0& - 9.45 & 0.08 \\
UGC08508 &   63.09&  17.91 & 2.69 .14&  2.67 &   56  2&  181 & -0.8 &  M81&  3 &  10&  24.3& -13.1& -10.20 &-0.25 \\
DDO181   &   78.09&  18.58 & 3.01 .15&  3.08 &  214  2&  284 & -1.1 &N4736&  68 &  10&  24.7& -13.2& - 9.78 &-0.20 \\
DDO183   &   81.14&  20.47 & 3.22 .16&  3.38 &  188  2&  254 & -1.0 &N4736&  13 &  10&  24.8& -13.2& -10.06 &-0.21 \\
KKH86    &  116.34&  15.47 & 2.59 .13&  2.89 &  287  1&  209 & -1.4 &N5128&  10 &  10&  25.1& -10.3& -10.69 &-0.60 \\
UGC08833 &   83.54&  21.09 & 3.08 .15&  3.24 &  221  2&  280 & -1.1 &N4736&  51 &  10&  24.2& -12.2& -10.26 &-0.37 \\
KK230    &   84.56&  23.55 & 2.14 .10&  2.26 &   63  2&  127 & -1.3 &  M81& -12 &  10&  25.5& - 9.2& - 9.96 & 0.17 \\
DDO187   &   97.83&  24.35 & 2.24 .12&  2.43 &  160  2&  180 & -1.4 &  MW &  24 &  10&  24.1& -12.5& -10.04 & 0.04 \\
DDO190   &   74.07&  26.85 & 2.80 .14&  2.84 &  150  2&  263 & -1.2 &  M81&  69 &   9&  23.6& -14.2& -10.17 &-0.40 \\
KKR25    &   56.09&  40.37 & 1.86 .12&  1.79 &  -79  8&  128 & -1.0 &  M31&  37 &  -1&  25.8& - 9.4& -11.82 &-1.80 \\
IC4662   &  199.19&   8.61 & 2.44 .18&  2.74 &  302  3&  139 & -1.3 &N5128& -46 &   9&  22.4& -15.5& -10.10 &-0.49 \\
SagdIr   &  221.27&  55.51 & 1.04 .07&  1.14 &  -79  1&   21 & -0.4 &  MW &  1 &  10&  24.8& -11.5& - 9.65 & 0.43 \\
DDO210   &  252.08&  50.24 & 0.98 .05&  0.97 & -140  2&   11 & -0.3 &  MW &   11 &  10&  24.6& -11.1& -10.64 &-0.33 \\
IC5152   &  234.23&  11.53 & 1.97 .12&  2.08 &  122  2&   73 & -1.3 & N253& -48 &   9&  23.8& -15.6& -10.27 &-0.70 \\
KK258    &  255.48&  18.58 & 2.23 .11&  2.34 &   92  5&  150 & -1.1 & N253&  3 &  -3&  26.0& -10.5& -11.64 &-1.41 \\
Tucana   &  227.61&  -0.92 & 0.88 .04&  1.09 &  194  4&   73 & -0.2 &  MW &  59 &  -1&  26.5& - 9.2& -12.66 &-1.17 \\
UGCA438  &  258.88&   9.28 & 2.18 .12&  2.11 &   62  8&   99 & -0.4 &  N55& -25 &  10&  24.1& -12.9& -10.24 &-0.34 \\
KKH98    &  332.35&  23.17 & 2.52 .13&  2.11 & -132  2&  156 & -0.9 &Maffei2&  32 &  10&  25.1& -10.8& - 9.90 &-0.13 \\
\hline
\end{tabular}
 %\end{rotate}
%\end{center}
\end{table}
 %\end{rotate}

\clearpage
\begin{figure*}
\includegraphics[scale=0.6,angle=-90]{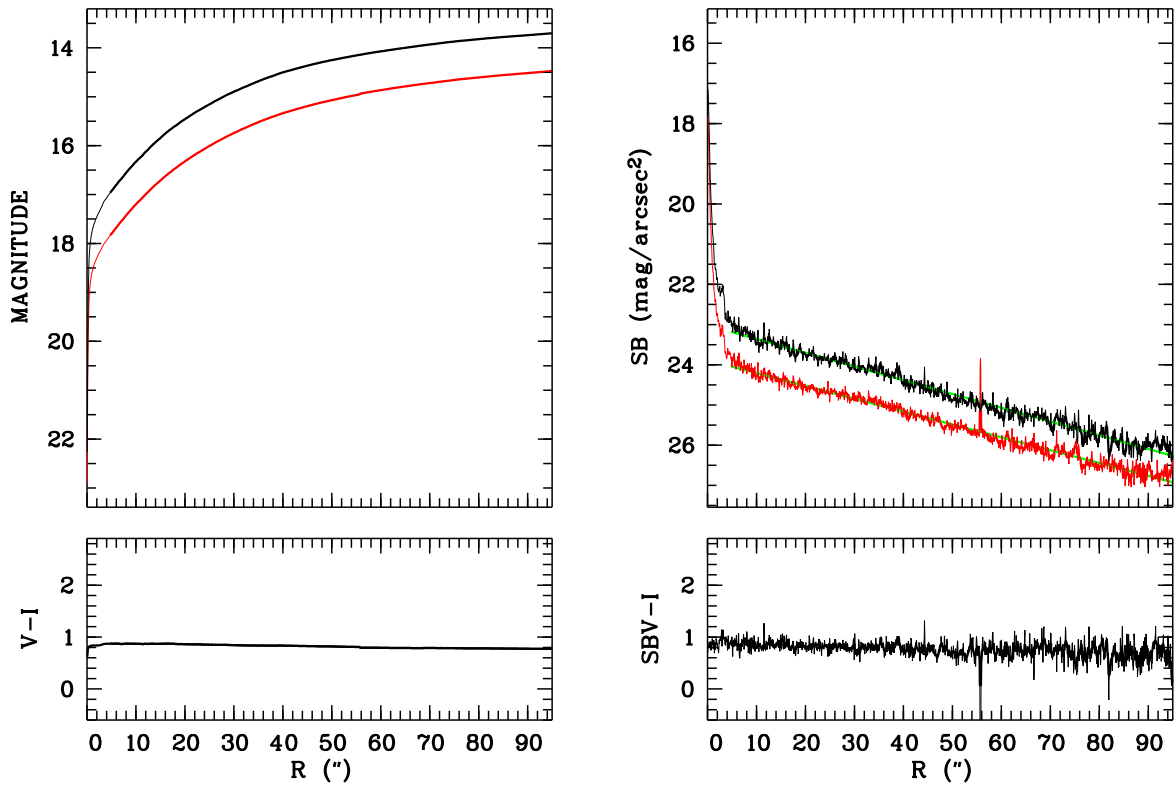}
\caption{ Azimuthally averaged surface brightness profiles of KKS3
obtained with the HST/ACS. The I-band profile is going above the V-band one.}
\end{figure*}

\begin{figure*}
\includegraphics[scale=0.6,angle=-90]{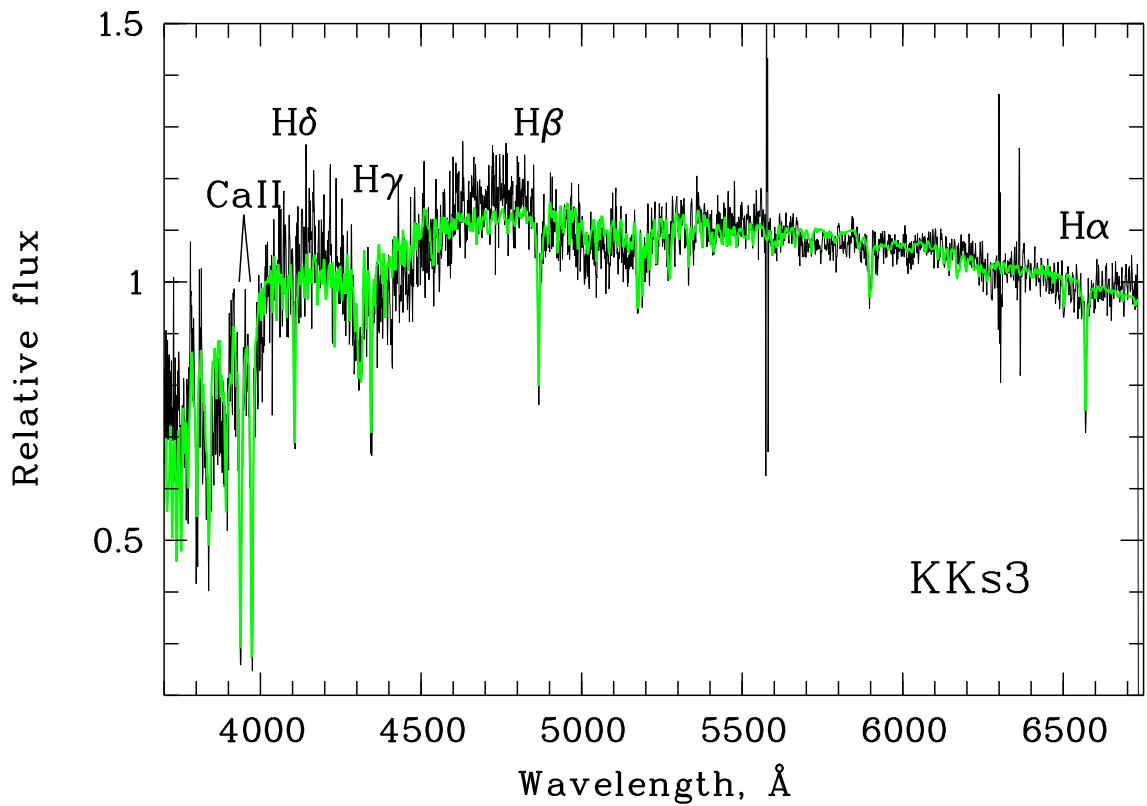}
\caption{ One-dimensional scan of the spectrum of the globular
cluster in KKs3 obtained with the SALT telescope. The fitting spectra
was carried out with a composite model (green line) using the
standard ULySS package }
\end{figure*}

\begin{figure*}
\includegraphics[scale=0.6,angle=-90]{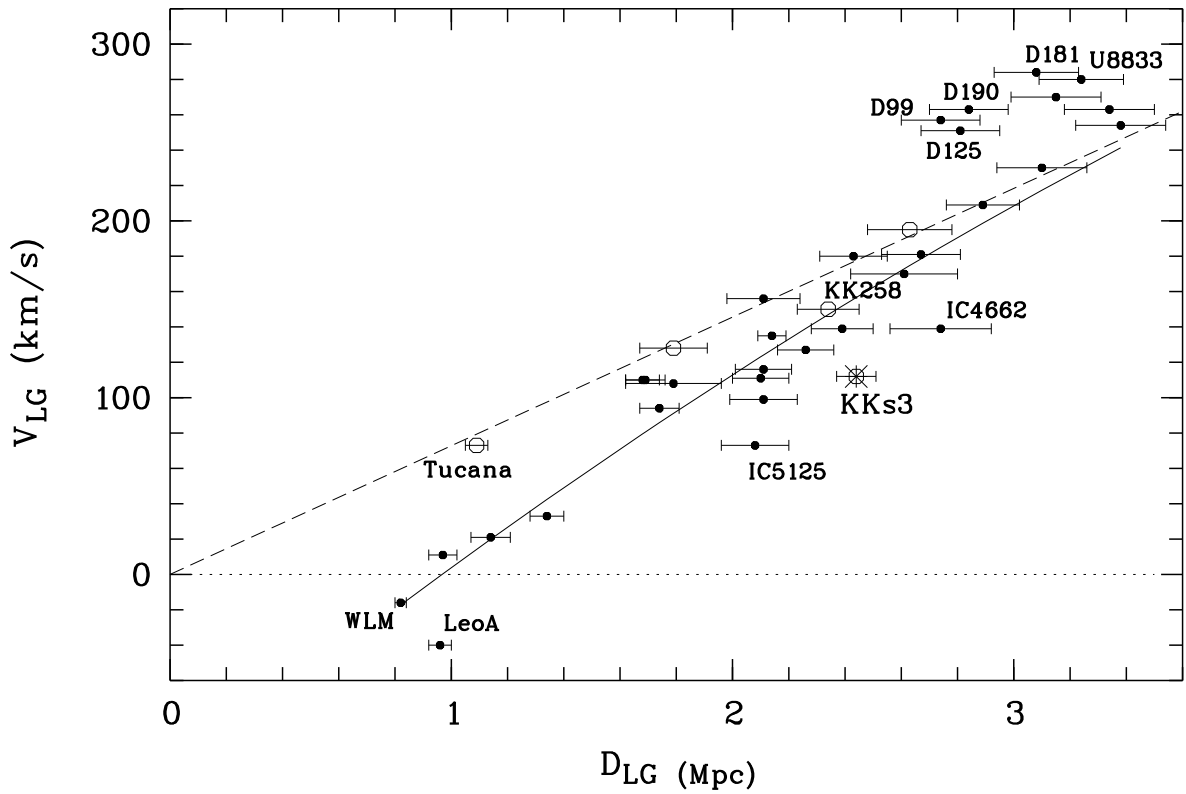}
    \caption{The Hubble diagram for the nearby field galaxies.
     The galaxies of 
late $(T>0$) and early types are denoted by the solid and open
circles respectively (with distance errors). The dashed straight
line denotes the undisturbed Hubble flow with a parameter of $H_0$
=73 km s$^{-1}$ Mpc$^{-1}$. The solid line corresponds to the
regression line $\langle V_{LG}|D_{LG}\rangle$ which crosses the
zero-velocity line at $R_0=0.96$ Mpc.}  
    
\end{figure*}

\begin{figure*}
\includegraphics[scale=0.6,angle=-90]{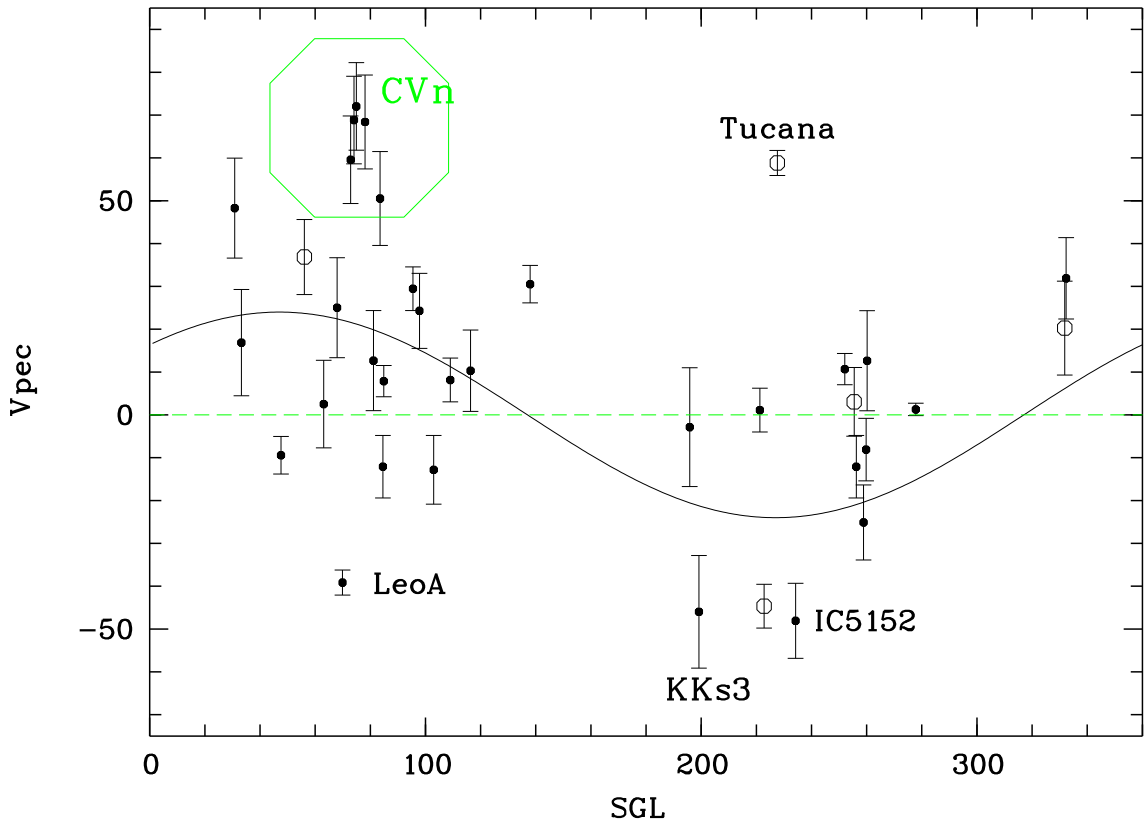}
     \caption{Peculiar velocities of the nearby isolated galaxies in km s$^{-1}$ as 
    a function of their supergalactic longitude. The
    wavy line corresponds to the dipole component $V_d = 24 \cos$ (SGL--47).
    The notations for the late and early type galaxies are the same as in Fig.~3.
    The vertical intervals denote the errors in estimation of the galaxy 
    velocities caused by the distance errors.}
\end{figure*}

\begin{figure*}
\includegraphics[scale=0.6,angle=-90]{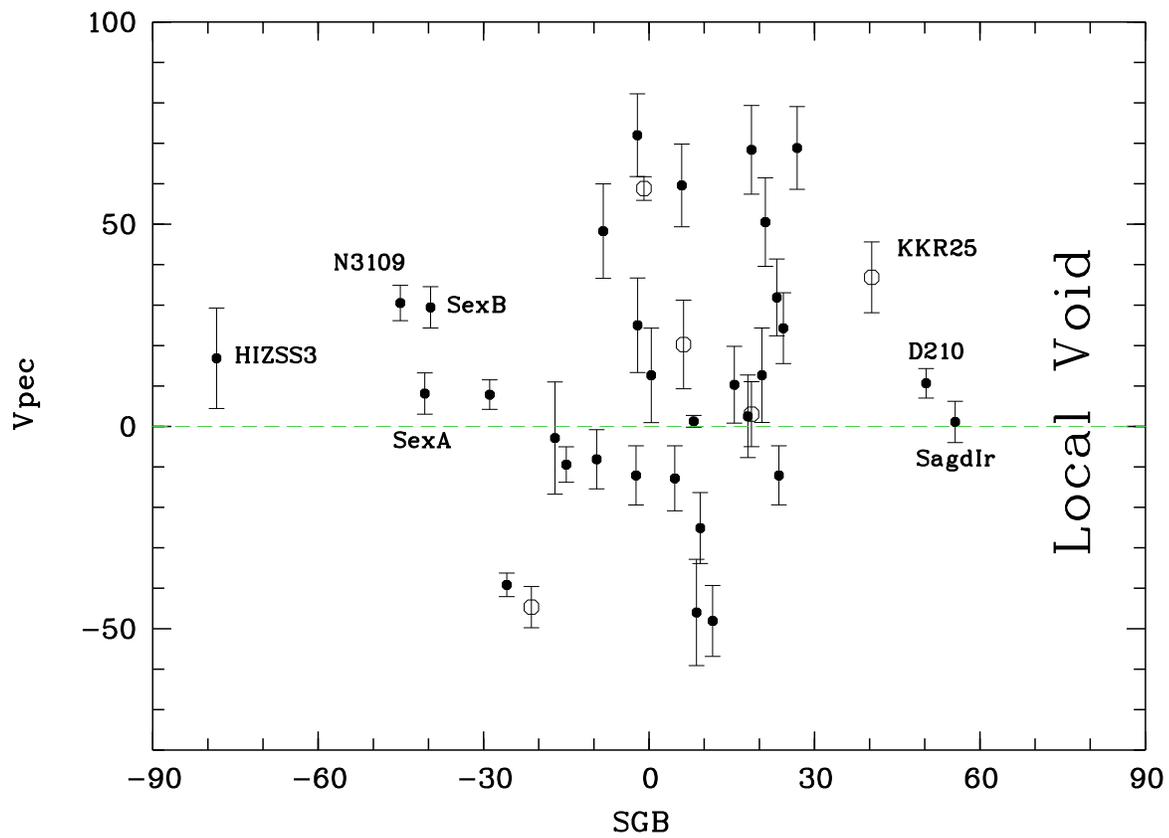}
     \caption{Peculiar velocities of the nearby field galaxies in km s$^{-1}$
 versus their supergalactic latitude. The notations for the late and early type galaxies are the same as in Fig.3.}
\end{figure*}

\begin{figure*}
\includegraphics[scale=0.3,angle=-90]{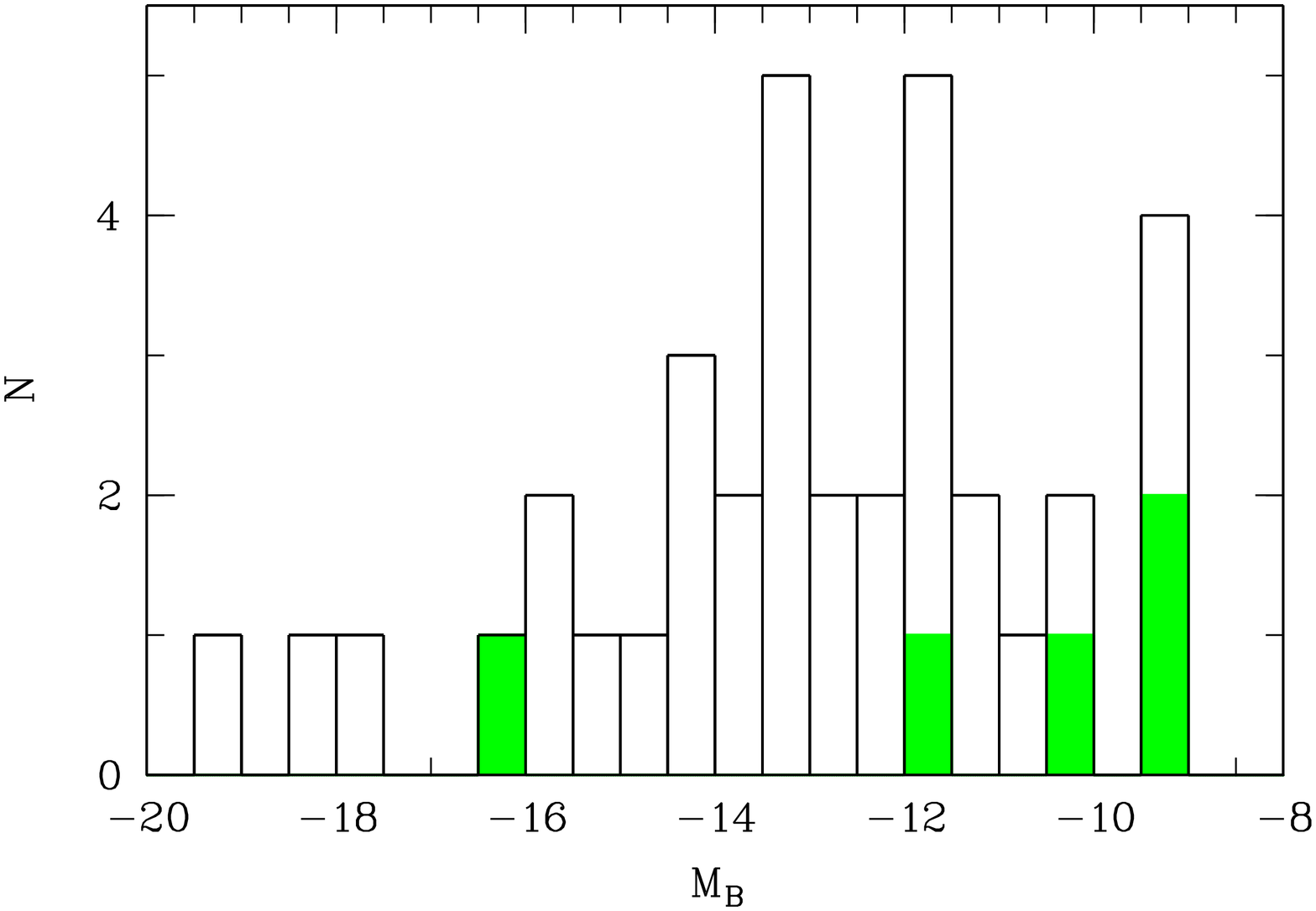}
    \includegraphics[scale=0.3,angle=-90]{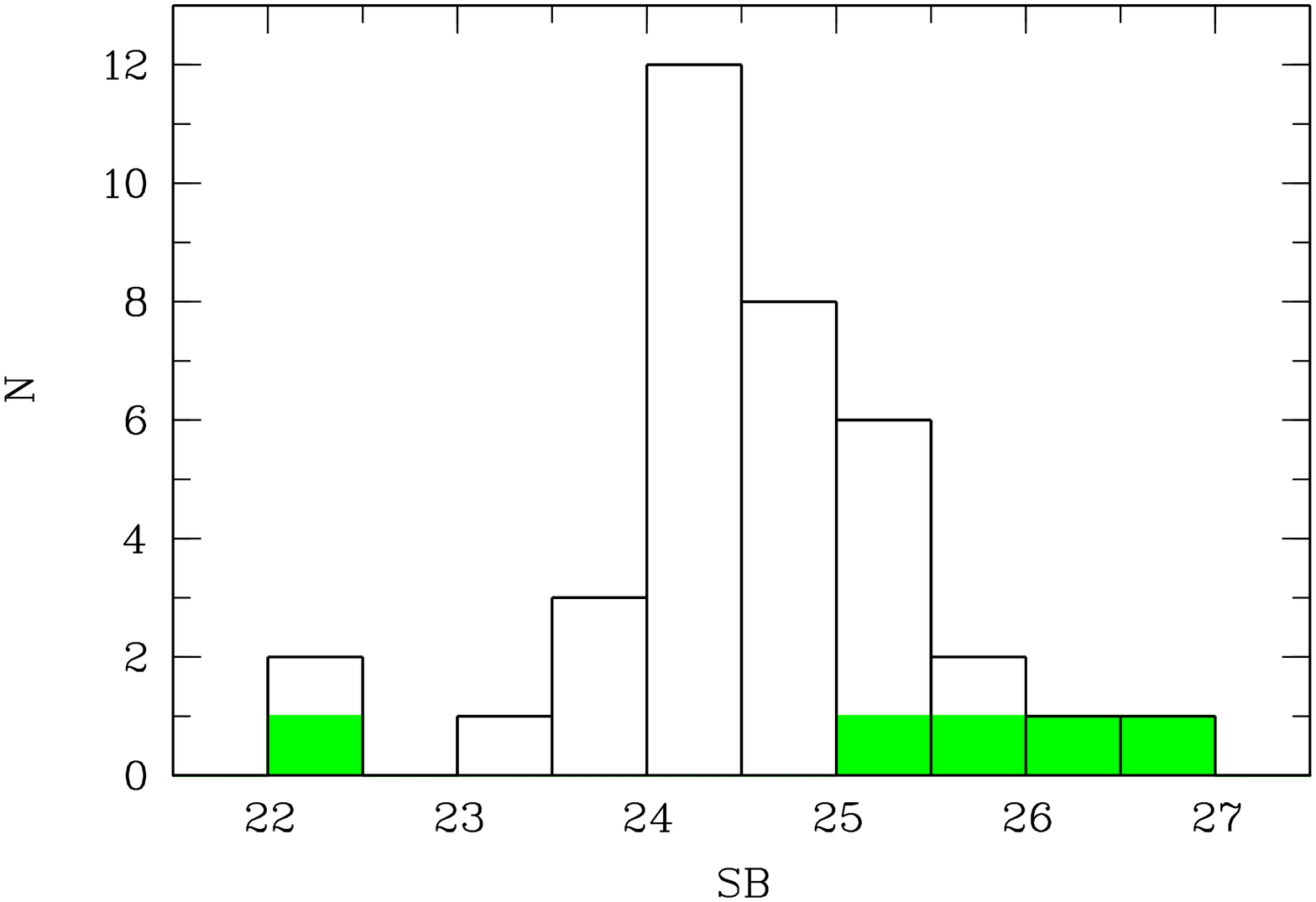}\\
    \includegraphics[scale=0.3,angle=-90]{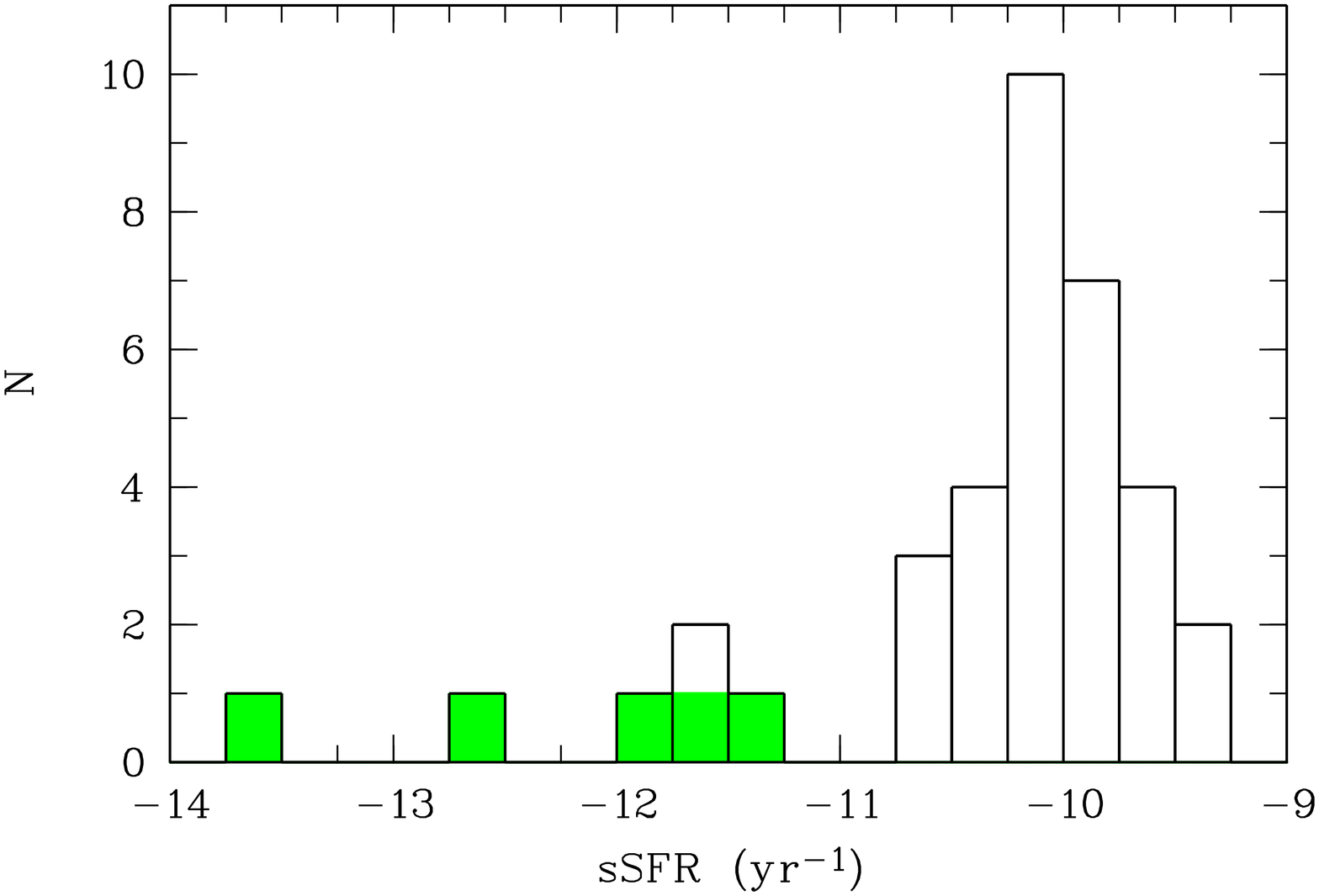}
    \includegraphics[scale=0.3,angle=-90]{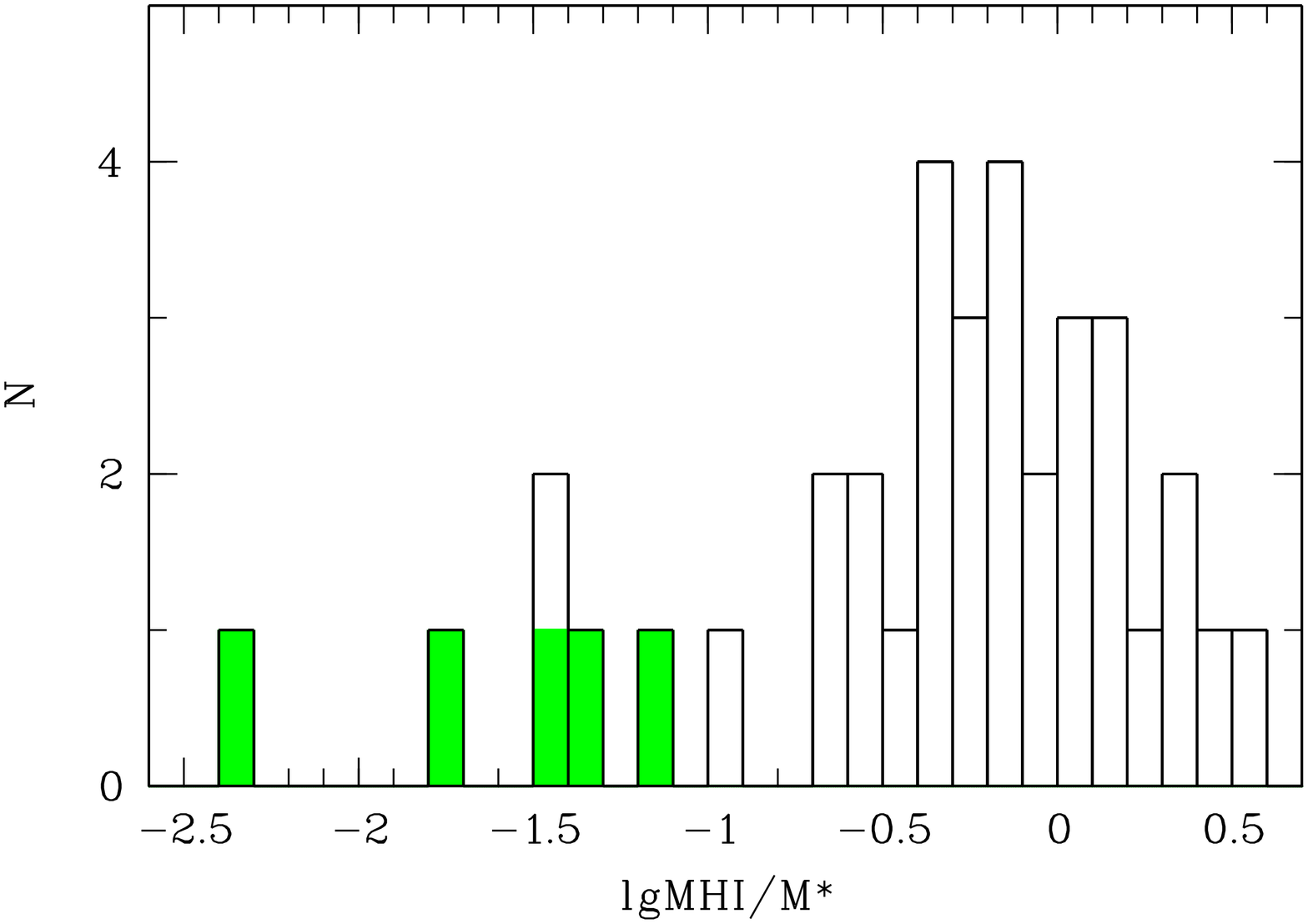}
 \caption{A mosaic of histograms showing the distribution of 36 nearest field galaxies according to different parameters:
    the absolute B-magnitude, the average surface brightness within the Holmberg radius, the specific star formation rate, and the hydrogen mass-to-
    stellar mass ratio. The early type (S0, dSph, dTr) galaxies are highlighted in green.}
    \end{figure*}


\begin{thebibliography}{}

 \bibitem{} Bell, E.F., McIntosh, D.H., Katz, N., \& Weinberg, M.D., 2003, ApJS, 149, 289

\bibitem{} Ben\'{i}tez-Llambay A., Navarro J.F., Abadi M.G., Gottl\"{o}ber S.,
   Yepes G., Hoffman Y., Steinmetz M., 2013, ApJ, 763, L41

\bibitem{} Buckley, D.A.H., Swart, G.P., \& Meiring, J.G., 2006,
  in Society of Photo-Optical Instrumentation Engineers
  (SPIE) Conference Series Vol. 6267 of Society of Photo-
  Optical Instrumentation Engineers (SPIE) Conference
  Series, Completion and commissioning of the Southern
  African Large Telescope

\bibitem{} Burgh, E. B., Nordsieck, K. H., Kobulnicky, H. A. et al., 2003,
  in Iye M., Moorwood A. F. M., eds, Instrument Design
  and Performance for Optical/Infrared Ground-based Telescopes
  Vol. 4841 of Society of Photo-Optical Instrumentation
  Engineers (SPIE) Conference Series, Prime Focus
  Imaging Spectrograph for the Southern African Large
  Telescope: optical design. p. 1463

\bibitem{} Crawford, S. M., Still, M., Schellart, P. et al., 2010, in Society of Photo-Optical Instrumentation
  Engineers (SPIE) Conference Series Vol. 7737 of Society
  of Photo-Optical Instrumentation Engineers (SPIE)
  Conference Series, PySALT: the SALT science pipeline

\bibitem{} Chernin, A.D., Emelyanov, N.V. \& Karachentsev, I.D., 2015, MNRAS,
449, 2069

\bibitem{} Chernin, A.D., 2008, UFN, 178, 267

\bibitem{} Chernin, A.D., Karachentsev, I.D., Valtonen, M.I., et al., 2004,
A\&A, 415, 19

\bibitem{} Chernin, A.D., 2001, Physics-Uspekhi, 44, 1099

\bibitem{} Ekholm, T., Baryshev, Y., Teerikorpi, P. et al., 2001, A \& A, 368,
L17

\bibitem{} Fraternali, F., Tolstoy, E., Irwin, M.J., \& Cole, A.A., 2009, A \& A,
499, 121

\bibitem{} Gil de Paz, A., et al., 2007, ApJS, 173, 185

\bibitem{} Giovanelli, R., Haynes, M.P., Adams, E.A.K., et al., 2013, AJ, 146, 15

\bibitem{} Hoffman, Y., Martinez-Vaquero, L.A., Yepes, G., \& Gottloeber, S., 2008,
MNRAS, 386, 390

\bibitem{} Huchtmeier, W.K., Karachentsev, I.D., \& Karachentseva, V.E., 2000, A\&A,
147, 187

\bibitem{} Karachentsev, I.D., Makarova, L.N., Makarov, D.I., Tully, R.B., \&
Rizzi, L., 2015, MNRAS, 447L, 85

\bibitem{} Karachentsev, I.D., Makarova, L.N., Tully, R. B., Wu, P.-F., \& Kniazev
A.Y., 2014, MNRAS, 443, 1281

\bibitem{} Karachentsev, I.D., Makarov, D.I., \& Kaisina, E.I., 2013, AJ, 145,
101 (=UNGC)

\bibitem{} Karachentsev, I.D., Kashibadze,O.G., Makarov, D.I., \& Tully, R.B.,
2009, MNRAS, 393, 1265

\bibitem{} Karachentsev, I.D., 2005, AJ, 129. 178

\bibitem{} Karachentsev, I.D., Sharina, M.E., Makarov, D.I., et al. 2002, A\&A,
389, 812

\bibitem{} Karachentsev, I.D., \& Makarov, D.I., 2001, Astrofizics, 44, 5 

\bibitem{} Karachentsev,I.D., Makarov, D.I., 1996, AJ, 111, 794

\bibitem{} Kniazev, A.Y., Zijlstra, A.A., Grebel, E.K., et al. 2008, MNRAS,
388, 1667

\bibitem{} Kobulnicky, H.A., Nordsieck, K.H., Burgh, E.B., et al., eds, Instrument Design
  and Performance for Optical/Infrared Ground-based
  Telescopes Vol. 4841 of Society of Photo-Optical Instrumentation
  Engineers (SPIE) Conference Series, Prime focus
  imaging spectrograph for the Southern African large
  telescope: operational modes. p. 1634

\bibitem{} Makarov, D.I., Makarova, L.N., Sharina, M.E., et al. 2012,
MNRAS, 425, 709

\bibitem{} O'Donoghue, D., Buckley, D.A.H., Balona, L.A., et al., 2006, MNRAS, 372, 151

\bibitem{} Oosterloo, T., Da Costa, G.S., \& Staveley-Smith, L., 1996, AJ, 112,
1969

\bibitem{} Peirani, S., \& de Freitas Pacheco,J.A. 2008, A\&A, 488, 845

\bibitem{} Rizzi, L., Tully, R.B., Makarov, D., et al., 2007, ApJ, 661, 815

\bibitem{} Sirianni, M., Jee, M. J., Bentez, N., et al., 2005, PASP, 117, 1049

\bibitem{} Shaya, E.J., Peebles, P.J.E., \& Tully, R.B., 1995, ApJ, 454, 15

\bibitem{} Teerikorpi, P., Chernin, A.D., \& Baryshev, Y.V., 2005, A\&A, 440, 791

\bibitem{} Tikhonov, A.V., \& Klypin, A., 2009, MNRAS, 395, 1915

\bibitem{} Tully, R.B., Rizzi, L., Shaya, E.J., et al., 2009, AJ, 138, 332 (=EDD)

\bibitem{} Tully, R.B., Shaya, E.J., Karachentsev, I.D., et al., 2008, ApJ, 676,
184

\bibitem{} van de Weygaert, R., \& Schaap, W. 2007, in Data Analysis in
Cosmology, ed. Martinez,  V., et al. (Berlin: Springer)

\end{thebibliography}
\end{document}